\documentclass[conference,compsoc]{IEEEtran}

\usepackage{times}
\usepackage{cite}
\usepackage{url}
\usepackage{xspace}
\usepackage{booktabs}
\usepackage{multirow}
\usepackage{array}
\usepackage{enumitem}
\usepackage{amsmath,amssymb}
\usepackage{graphicx}
\usepackage[table]{xcolor}
\usepackage[caption=false,font=footnotesize]{subfig}
\usepackage{balance}
\usepackage{color}
\usepackage{hyperref}
\usepackage{pifont}
\usepackage{tabularx}
\usepackage{makecell}
\usepackage{tikz}
\usetikzlibrary{arrows.meta,positioning}
\usepackage[most]{tcolorbox}
\newcolumntype{Y}{>{\raggedright\arraybackslash}X}

\newcommand{\tool}{\textsc{Kauge}\xspace} 
\newcommand{\scp}{SCP\xspace}
\newcommand{\scps}{SCPs\xspace}

\newcommand{\llms}{LLMs\xspace}

\newcommand{\passatk}{\textsf{pass@}k\xspace}

\newcommand{\securepassatk}{\textsf{secure-pass@}k\xspace}
\newcommand{\seccondk}{\textsf{secure@pass}k\xspace}  

\newcommand{\newpara}[1]{\smallskip\noindent\textbf{#1.}}

\definecolor{SOKBlue}{HTML}{4477AA}
\definecolor{SOKRed}{HTML}{CC6677}
\definecolor{SOKGreen}{HTML}{228833}
\definecolor{SOKPurple}{HTML}{AA3377}
\definecolor{SOKGray}{HTML}{666666}
\definecolor{SOKLightBlue}{HTML}{8DB9D8}
\definecolor{SOKLightGreen}{HTML}{A9D18E}
\definecolor{SOKGold}{HTML}{F2C879}
\definecolor{SOKPink}{HTML}{D9A0A8}
\definecolor{SOKLightGray}{HTML}{C9C9C9}
\definecolor{SOKLightPurple}{HTML}{B9A4D8}
\newcommand{\legendbox}[1]{\colorbox{#1}{\rule[-1pt]{0pt}{5.5pt}\hspace{0.6em}}}
\newcommand{\legendline}[1]{\textcolor{#1}{\rule[2pt]{0.8em}{1.5pt}}}
\newcommand{\stat}[1]{\textcolor{SOKPurple}{\textbf{#1}}}

\newcounter{takeawaycnt}
\newtcolorbox{takeaway}{enhanced, breakable, colback=black!4, colframe=black!55,
  boxrule=0.6pt, arc=2pt, left=5pt, right=5pt, top=4pt, bottom=4pt, boxsep=2pt,
  before upper={\refstepcounter{takeawaycnt}\textbf{Takeaway~\thetakeawaycnt.}\space}}

\title{
SoK: AI Secure Code Generation: Progress, Pitfalls, and Paths Forward
}

\author{
\IEEEauthorblockN{Rupam Patir, Keyan Guo, Haipeng Cai, and Hongxin Hu}
\IEEEauthorblockA{University at Buffalo, SUNY\\
\{rupampat, keyanguo, haipengc, hongxinh\}@buffalo.edu}
}

\begin{document}
\maketitle

\begin{abstract}

The increasing use of AI systems for code generation raises a central security question: what can today's models and coding agents actually do to produce secure code, where do they still fail, and what would move the field forward? Existing work has explored prompting, fine-tuning, reinforcement learning, and agentic workflows for secure code generation, but the field still lacks a systematic understanding of how these techniques improve security and why substantial failures persist. In this SoK, we systematize the progress, pitfalls, and paths forward for AI secure code generation. We introduce a three-level framework that measures models' natural-language understanding of secure coding principles, their code-level actuation of those principles during generation, and the knowledge--actuation gaps between the two. We instantiate this framework across models and coding agents on benchmarks covering both isolated function-level security and full web-application security. Our results show that secure-coding-principle understanding is a statistically strong predictor of code-level outcomes, including functional correctness, security, and joint functional-security correctness. Yet substantial knowledge--actuation gaps remain: models can recognize relevant security principles but still fail to translate them into secure and functional code. These findings offer a principle-centered account of where AI secure code generation stands today and identify concrete paths forward through principle-guided generation, evaluation, benchmarking, and agentic workflows.
\end{abstract}


\section{Introduction}
\label{sec:intro}


Large language models (\llms) and coding agents now write, complete, review,
and integrate software at
scale~\cite{pearce2022asleep,perry2023users,sandoval2023lostatc}.
Their adoption changes the security question from a relatively narrow one---``can a model
produce a vulnerable snippet?"---to a broader systems question: 
``\emph{what progress has the field of AI secure code generation made, what pitfalls still prevent reliable security, and what paths forward would make generated code safer in practice?}"

The field has made real progress. Early empirical studies showed that code
assistants often emit vulnerable code and can lead developers to write less
secure programs~\cite{pearce2022asleep,perry2023users,khoury2023chatgpt}.
Since then, researchers have proposed security-aware prompting~\cite{tony2025prompting,pearce2023repair},
retrieval-augmented generation~\cite{shi2025rescue,zhang2024seccoder},
constrained decoding~\cite{fu2024constrained,scodegen2025}, supervised
fine-tuning~\cite{sven2023,safecoder2024}, reinforcement learning and
proactive alignment~\cite{xu2024prosec,liu2026purpcode,hasan2025teaching},
and agentic scaffolds that let models access files, shells, tests, and
development environments~\cite{nunez2024autosafecoder,saul2025scgagent,yang2024sweagent}.
Benchmarks have also improved, moving from static or pattern-based snippet
datasets~\cite{siddiq2022securityeval,tony2023llmseceval,bhatt2023cyberseceval,bhatt2024cyberseceval2,hajipour2024codelmsec,codesecEval2024}
to executable function-level and application-level
suites~\cite{secodeplt2025,cweval,baxbench}, and more recently to repository
and agent-oriented evaluations~\cite{secrepobench2025,realsecbench2026,secbench2025}.
These studies show that interventions can change generated code and that
stronger systems can sometimes produce code that is both functional and secure
under benchmark oracles.

What remains missing, however, is a mechanism-level account of this progress: \emph{what} these
interventions actually changed and \emph{why} substantial failures persist. 
Most evaluations reduce secure generation to an output verdict: \textit{did the program pass
functional tests, trigger a vulnerability detector, or block an
exploit}~\cite{dai2025comprehensive,dai2026rethinking,chen2024slr}? Such
verdicts are necessary, but they do not explain \emph{why} security succeeds or
fails. A model may not know the relevant security practice; it may know the
practice but fail to recognize that it applies; it may apply a plausible but
incomplete mitigation; it may place a check at the wrong boundary; or it may
block an exploit only by breaking required functionality. These cases demand
different remedies, yet conventional metrics often collapse them into the same
pass/fail outcome. 
In particular, static-analysis-based evaluations compound this: tool coverage varies by
language, framework, and rule set, and a clean verdict can mask a genuinely
exploitable flaw that a dynamic or exploit-based oracle would
catch~\cite{dai2026rethinking,cweval}.
Distinguishing these failure modes requires a lens that current evaluation
practices do not provide.


In this SoK, we propose that secure coding principles (\scps) provide an effective mechanism 
for separating these cases. Established catalogs such as OWASP~\cite{owasp_scp} and
CERT~\cite{cert_c} express security knowledge as prescriptive rules: e.g., ``validate
attacker-controlled inputs", ``use parameterized queries", ``canonicalize paths before
file access", ``avoid weak cryptographic primitives", and ``enforce authorization at
protected boundaries". Unlike CWE labels, which name what failed after the fact,
\scps describe what correct generation should do~\cite{owasp_scp,realsecbench2026}. 
Crucially for a study of \emph{language} models, 
\scps are also written in
natural language, the \textit{native modality} in which \llms are prompted, trained, and
evaluated. 
An \scp is a natural-language instruction the model can, or fail to, read, reason about, and follow~\cite{tony2025prompting,safecoder2024}. 
Moreover, compared to a CWE identifier that is an opaque label, 
\scps are more granular: a single principle such as input validation
spans many CWEs, while a single CWE such as command injection may require
several principles together~\cite{owasp_scp,cweval}. 
Thus, organizing measurement around \scps lets us ask not only
whether generated code is vulnerable, but whether the model understood the
relevant practice and implemented the intended defensive mechanism.


With these insights, we introduce \tool
(\underline{K}nowledge--\underline{A}ctuation
\underline{U}nified \underline{G}ap \underline{E}valuation), a
\textit{principle-centered framework for systematizing AI secure code generation}.%
\footnote{The code is available at
\url{https://github.com/rupampatir/SoK_KAUGE}, and the benchmark at
\url{https://doi.org/10.5281/zenodo.20820512}.}
\tool decomposes AI secure code generation into \textbf{three layers}. Layer~1, \emph{Knowledge},
measures whether models understand \scps as natural-language concepts: what a
practice states, why violating it creates a vulnerability, how it should be
applied, and when it is relevant. Layer~2, \emph{Actuation}, measures whether
models translate the knowledge/understanding to generating code that is both functional and exploit-resistant under
executable benchmark oracles. Layer~3, \emph{Gap}, examines why Layers 1 and 2 diverge by checking whether generated
programs implement the exploit-specific defensive mechanism suggested by the
relevant \scps. Together, these layers separate three questions that existing
benchmarks often conflate: \textit{what a system knows}, \textit{what it produces}, and \textit{whether
the produced code actuates the intended security practice}. 
Across the three layers, \tool turns comparison/benchmarking into 
explanatory diagnosis.


We instantiate \tool with three constructed artifacts: a natural-language-understanding (NLU) benchmark over
OWASP and CERT principles, an exploit-to-SCP defense formula mapping, and an
SCP-compliance judge. The NLU benchmark contains 6,382 verified fixed-answer
questions generated from 456 source rules and organized across declarative,
causal, procedural, and contextual reasoning dimensions. The defense mapping
links executable exploits to alternative sufficient \scp strategies, and the compliance judge checks whether the generated code implements those strategies. We use these artifacts to study models, prompting methods, security fine-tuning, and reinforcement-learning-based post-training, and agentic scaffolding on two executable benchmarks:
CWEval~\cite{cweval} for isolated 
function-level security and BaxBench~\cite{baxbench} for full
web-application security.

Our findings show both progress and a persistent pitfall. At the \textit{Knowledge}
layer, current systems already understand secure coding practices well: scores
are far above chance across OWASP and CERT catalogs, with the clearest remaining
weakness in causal reasoning over lower-level CERT C/C++ rules. At the
\textit{Actuation} layer, however, secure generation remains much harder. Function-level
failures split between code that does not meet the specification and code that
works but remains exploitable. At application scale, functionality itself often
becomes the binding constraint: many systems fail to produce a runnable backend
before security can be meaningfully tested. 

The central result is that knowledge and actuation are related but not
equivalent. Natural-language \scp understanding is strongly statistically correlated 
with deployable secure generation, including functional correctness, security,
and joint functional-security correctness. However, actuation is systematically
lower than knowledge. Models often know the relevant principle but fail to turn
it into the right program behavior at the right implementation boundary. This
is the knowledge--actuation gap.


The \textit{Gap} layer explains what ordinary metrics hide. Some outputs are
\emph{principled successes}: they implement the mapped defense and block the
exploit. Others are \emph{secure by other means}: they block the exploit through
a mechanism outside the mapped formula. Conversely, some outputs are
\emph{compliant but vulnerable}: they appear to implement the relevant principle
but leave the exploit open because the abstract principle underdetermines the
sink-specific implementation. Finally, some outputs are direct
\emph{actuation failures}: the program is functional and exploitable, with no
evidence that the relevant security principles (practice) were applied. These cases also show why secure
code generation cannot be understood through pass/fail metrics alone. 

These diagnoses also enlighten \textit{paths forward}. 
Merely exposing models to more secure-coding principles is not enough. 
Secure code generation needs to optimize the translation from security principle to
secure program: executable feedback should preserve functionality while blocking
exploits; principle guidance should become sink-aware and
implementation-specific; training should connect attacker input, vulnerable
sink, applicable principle, and minimal secure code change; and benchmarks
should report not only whether code is functional and secure, but also whether
it implements the intended defensive mechanism.


In summary, this paper makes following contributions. 
\begin{itemize}[noitemsep,leftmargin=*,topsep=2pt]
    \item We \textbf{systematize progress} in AI secure code generation,
    organizing models, prompting/training methods, agents, and
    benchmarks by where they intervene in the generation pipeline and what security knowledge they rely on (\S\ref{sec:systematization}).

    \item We introduce \textbf{\tool}, a principle-centered measurement
    framework comprising three constructed artifacts: an NLU benchmark for
    secure-coding knowledge, an exploit-to-SCP defense formula mapping, and an
    SCP-compliance judge. 
    These artifacts separate secure-coding
    knowledge, code-level actuation, and the gap between them 
    (\S\ref{sec:kauge}--\ref{sec:layer3}).

    \item We show that natural-language \scp understanding is
    \textbf{strongly correlated with code-level generation outcomes} statistically,
    including functional correctness, security, and joint functional-security
    correctness, while substantial knowledge--actuation gaps remain
    (\S\ref{sec:layer1}, \S\ref{sec:layer2}, \S\ref{sec:layer3}).

    \item We \textbf{diagnose pitfalls} that existing evaluation output metrics hide, including
    functionality-security tradeoffs, weak causal understanding,
    secure-by-other-means behavior, compliant-but-vulnerable behavior, and
    direct actuation failures where known principles are not applied
    (\S\ref{sec:layer3}).

    \item We \textbf{identify paths forward} in AI secure code generation:
    executable feedback, sink-aware principle guidance,
    functionality-preserving generation, causal-actuation training,
    whole-project agent evaluation, and mechanism-aware evaluation and reporting
    (\S\ref{sec:paths}).
\end{itemize}

\section{Systematizing Progress}
\label{sec:systematization}

\subsection{Methodology}
\label{sec:sok-method}


\begin{table*}[t]
\centering
\caption{Taxonomy of secure-code-generation techniques.}
\label{tab:taxonomy}
\small
\setlength{\tabcolsep}{3pt}
\renewcommand{\arraystretch}{1.22}
\begin{tabularx}{.95\textwidth}{@{}p{4cm}Y Y Y}
\toprule
\textbf{Family (Intervention point)} & \textbf{Mechanism} & \textbf{Security knowledge form} &
\textbf{Representative studies} \\
\midrule

Input \& context steering \newline \emph{(pre-generation)} &
Supplies security-relevant context before generation through prompt guidance or
retrieved evidence. &
Prompt hints, demonstrations, self-critique instructions; retrieved rules and
examples. &
SecCoder~\cite{zhang2024seccoder}; RESCUE~\cite{shi2025rescue}; prompting
techniques~\cite{tony2025prompting} \\
\addlinespace[0.25em]

Training \& alignment \newline \emph{(model post-training)} &
Fine-tunes, instruction-tunes, or preference/RL-aligns the model so secure
coding behavior is internalized. &
Secure/insecure pairs; vulnerable--fixed pairs; synthetic repairs; preference or
RL rewards. &
SVEN~\cite{sven2023}; SafeCoder~\cite{safecoder2024};
HexaCoder~\cite{hajipour2024hexacoder}; LPO~\cite{hasan2025teaching};
ProSec~\cite{xu2024prosec}; PurpCode~\cite{liu2026purpcode} \\
\addlinespace[0.25em]

Constrained generation \newline \emph{(during generation)} &
Constrains decoding or candidate selection so insecure completions are
discouraged, filtered, or blocked without updating the model. &
Decoding-time specifications; constraint phrases; logit biases; automaton or
rule-based constraints. &
Constrained decoding~\cite{fu2024constrained}; SCodeGen~\cite{scodegen2025} \\
\addlinespace[0.25em]

Post-generation feedback \newline \emph{(post-generation / multi-step)} &
Treats generated seed code, checks it with external feedback, then refines it by repair or agentic iteration. &
Static-analysis warnings; fuzzing results; failing tests; exploit feedback;
secure-coding guidelines. &
PromSec~\cite{promsec2024}; AutoSafeCoder~\cite{nunez2024autosafecoder};
SCGAgent~\cite{saul2025scgagent} \\
\bottomrule
\end{tabularx}
\end{table*}

\newpara{Search sources}
We searched bibliographic databases that jointly cover the security, software-engineering, and machine-learning venues where this line of work appears: the ACM Digital Library, IEEE Xplore, USENIX, DBLP, and the ACL Anthology. Because a substantial fraction of recent work first circulates as preprints before formal publication, we also searched arXiv, focusing on cs.CR, cs.SE, and cs.LG, and cross-checked OpenReview when venue or acceptance status was relevant. The search window begins in 2021, when publicly available LLM-based code assistants such as OpenAI Codex and GitHub Copilot made the security of AI-generated code a practical concern, and extends through June 2026.

\newpara{Search procedure}
We constructed search terms along three axes, combined conjunctively: a model axis (large language models, code models, coding agents), a task axis (generation, completion, synthesis, repair), and a security axis (security, vulnerability, CWE, CVE, secure coding). We included common synonyms within each axis and adapted operators to each database.
We first screened the results returned by the keyword searches. We then applied backward and forward snowballing~\cite{wohlin2014snowballing} from the retained studies to identify papers that used different terminology but addressed the same problem space. After de-duplication, we screened the resulting corpus against the inclusion and exclusion criteria below. The final collection is summarized in Table~\ref{tab:taxonomy}.

\newpara{Inclusion criteria}
A study was included if it satisfied all of the following criteria:

\begin{itemize}
\item[(I1)] \textbf{Generative scope.} It uses an LLM, code-specialized model, or coding agent to generate, complete, repair, or transform code. Studies limited to detecting, localizing, or classifying vulnerabilities in existing code were included only if they also produced a concrete code fix or secure-generation intervention.

\item[(I2)] \textbf{Security objective.} It explicitly targets security, vulnerability reduction, secure coding, security-aware repair, or prevention of insecure generated code as a primary objective, rather than treating security as an incidental aspect of general code quality.

\item[(I3)] \textbf{Security-relevant evaluation.} It evaluates generated or repaired code using a security-relevant signal, such as vulnerability labels, CWE categories, static-analysis findings, dynamic or exploit-based tests, manual or expert audit, or established security benchmarks.

\item[(I4)] \textbf{Venue and verifiability.} It is either peer-reviewed at an archival venue or available as a preprint with sufficient methodological detail to assess the proposed technique, evaluation setting, and security claims. When applicable, we additionally considered the availability of artifacts, datasets, prompts, or evaluation scripts.
\end{itemize}

\noindent \textbf{Exclusion criteria.}
We excluded studies that: (E1) use LLMs solely for vulnerability detection, localization, or classification without any generative remediation or secure-generation component; (E2) focus on offensive security or LLM misuse, such as malware synthesis, exploit generation, or jailbreaks, without a defensive secure-code-generation objective; (E3) report only general code quality or functional correctness without a security-specific evaluation; (E4) are non-archival artifacts, such as blog posts, tool documentation, tutorials, or short abstracts lacking sufficient methodological detail; and (E5) are superseded by a later version from the same authors, in which case we retain the most complete version.


\subsection{Techniques Taxonomy}
\label{sec:taxonomy}



We organize secure-code-generation techniques into four families by where they intervene in the generation pipeline: \emph{input and context steering}, which supplies security knowledge before generation; \emph{training and alignment}, which internalizes security behavior in model parameters; \emph{constrained generation}, which enforces selected security requirements during decoding; and \emph{post-generation feedback}, which revises generated artifacts using external checks or agentic iteration. Table~\ref{tab:taxonomy} summarizes these families together with their mechanisms, security knowledge forms, and representative studies.

This intervention-point view matters because the form of security knowledge available to a method is coupled to when it acts: prompts and retrieved examples are available before generation, training pairs and rewards during model adaptation, constraints during decoding, and tests or exploit feedback only after code has been produced. 

\subsection{Techniques: From Knowledge to Code}
\label{sec:techniques}

We now examine each family in turn, reading its intervention point as a choice about where in the generation a method acts on knowledge and on the resulting code.

\newpara{Input level---supplying knowledge from outside the model}
The lowest-overhead interventions leave the model unchanged and place secure-coding knowledge in its context. The knowledge is either \emph{prompt-borne}---security
reminders, CWE hints, demonstrations, self-critique, multi-model collaboration---or \emph{retrieved} and injected as rules and examples. \textsc{SecCoder}~\cite{zhang2024seccoder} retrieves safe in-context demonstrations and stresses generalization to unseen vulnerabilities; \textsc{Rescue}~\cite{shi2025rescue} builds a security-specific retrieval pipeline over a distilled knowledge base; and systematic prompting studies~\cite{tony2025prompting} catalog strategies such as Recursive Criticism and Improvement. 
Being training-free and model-agnostic, this family is the most broadly deployable---but nothing in it guarantees the supplied knowledge is enacted: the model may receive the right principle and still not apply it. Tellingly, these methods are usually validated against static oracles, so even their reported gains may reflect the checker more than the code.

\newpara{Model level---internalizing knowledge into the weights}
A second family pushes knowledge into the model's parameters so that secure generation becomes the default rather than a prompted afterthought, and it is
dominated by the question of where the training signal comes from. The earliest signal is \emph{secure/insecure pairs}: \textsc{SVEN}~\cite{sven2023} learns property-specific prefix vectors from curated fixes, and \textsc{SafeCoder}~\cite{safecoder2024} folds security into ordinary instruction tuning. Because real vulnerable--fixed pairs are scarce, later work manufactures the signal---\textsc{HexaCoder}~\cite{hajipour2024hexacoder} synthesizes oracle-repaired pairs for LoRA tuning, \textsc{ProSec}~\cite{xu2024prosec} provokes vulnerabilities from CWE templates and learns fixes by preference optimization, and \textsc{LPO}~\cite{hasan2025teaching} distills preference pairs while masking security-relevant tokens so the signal is not washed out across
unchanged code. \textsc{PurpCode}~\cite{liu2026purpcode} instead internalizes explicit cybersafety \emph{rules} through reasoning-style reinforcement
learning. 
This family can produce more persistent changes than prompting because the intervention is built into the policy rather than supplied at inference time; but it depends on labeled data or a reward signal, trades against functional utility, generalizes unevenly across models and CWEs, and---for RL---can conflate safety refusal with secure coding.

\newpara{Decoding level---enforcing secure code directly}
A third family intervenes on the token distribution at generation time and is distinctive in our framing because it acts on the generated code directly,
largely independent of whether the model ``knows'' the principle: the secure behavior is a constraint mechanically enforced as the code is emitted.
\textsc{Constrained decoding}~\cite{fu2024constrained} shows that such constraints can outperform parameter-level steering with no specialized training data, while insisting that outputs be both secure \emph{and} correct (and contributing to the CodeGuard+ benchmark). \textsc{SCodeGen}~\cite{scodegen2025} makes the approach deployable, compiling constraints into a runtime automaton with logit modulation to preserve near-unconstrained latency under multiple concurrent constraints. The appeal is a hard guarantee without retraining; the limitation is that the constraint must be expressible at the token level, which bounds the vulnerability classes it can cover.

\newpara{Post-generation level---repairing code through feedback or agentic iteration}
The most recent family wraps the model in a generate--check--revise loop, turning an oracle's verdict into a new form of knowledge---a static-analysis warning, a fuzzing crash, a failing exploit---and feeding it back to repair the draft. \textsc{PromSec}~\cite{promsec2024} closes a single such loop, using a generative adversarial graph network to drive iterative prompt re-optimization until vulnerabilities clear. The agentic instances coordinate multiple steps:
\textsc{AutoSafeCoder}~\cite{nunez2024autosafecoder} composes coding, static-analysis, and fuzzing agents to add the dynamic runtime signal earlier families lack, while \textsc{SCGAgent}~\cite{saul2025scgagent} drives an agent with secure-coding guidelines and self-generated unit tests, letting a non-reasoning model rival dedicated reasoning models. Because they act after a first draft, these methods act directly on generated artifacts and can use execution feedback that earlier families lack; but they are computationally costly and bounded by tool and test coverage---an agent can satisfy the oracle while missing the underlying principle---and remain thinly evaluated at application scale.

\begin{takeaway}
Secure-code-generation methods differ in how directly they actuate security knowledge: prompts expose it, training internalizes it, constraints enforce it, and feedback or agents revise the generated artifact. The central bottleneck is therefore not only access to secure-coding knowledge, but functionality-preserving actuation in code.
\end{takeaway}

\subsection{Evaluation: What Benchmarks Can(not) See}
\label{sec:measurement}

If techniques act on knowledge and on the resulting code, the natural question is whether the field can \emph{see} either one. The answer is asymmetric: several years of benchmark progress have sharpened how we measure the code that comes out, while the knowledge that went in is almost never measured at all. Every reported gain is only as trustworthy as the oracle behind it, so we begin with how that oracle has evolved---and then with what it still cannot tell us.

\newpara{The measuring oracle}
The motivation was established early: empirical studies showed that code assistants emit vulnerable code at non-trivial rates and that developers using them can ship less secure software~\cite{pearce2022asleep, perry2023users,
sandoval2023lostatc, khoury2023chatgpt}. The benchmarks that followed differ along the oracle that decides whether code is secure. The first generation scored generation with \emph{static} signals---CWE-pattern matches, analyzer warnings, vulnerable/secure labels---over short, single-CWE prompts~\cite{siddiq2022securityeval,tony2023llmseceval, hajipour2024codelmsec,siddiq2024sallm,codesecEval2024}. A second wave scaled judging with rule- or LLM-based scoring~\cite{bhatt2023cyberseceval, llmseccode2024}. The most recent benchmarks \emph{execute}: they pair functional tests with concrete exploit or property tests, so ``secure'' means a specific attack provably fails against working code~\cite{secodeplt2025, cweval, baxbench}. The progression is consequential. A static label is a property of the \emph{checker}---it can credit code that is secure by other means, or silently miss an exploitable path---whereas a passed exploit test is a property of the \emph{code itself}. Yet this maturation, real as it is, runs entirely within the measurement of \emph{actuation}: every oracle in the progression scores the artifact a model emits, and a more faithful oracle simply scores that artifact more faithfully.

\begin{takeaway}
Security oracles have progressed from static label-matching to executable exploit tests, but most benchmarks still rely on static, single-CWE, function-level checks---so a large share of reported ``security'' is a property of the checker rather than of the code. And every step of this progress refines how the field scores the generated code, never what the model knew.
\end{takeaway}

\newpara{The measurement scope}
Even confined to the artifact, evaluation is uneven. The unit of judgment has grown from isolated functions---still the large majority of benchmarks---to repositories~\cite{secrepobench2025, secbench2025} and full applications~\cite{baxbench}, alongside a parallel line that targets repair of known-buggy programs~\cite{arvo2024, vulnrepaireval2025, drv2026}. But coverage clusters on a handful of CWE families and a few languages (Python-dominant), and
the setting closest to deployment---executable security conditioned on functional correctness---is represented by only a few recent benchmarks~\cite{secodeplt2025, cweval, baxbench}. We adopt two of these, CWEval (functions)~\cite{cweval} and BaxBench (web applications)~\cite{baxbench},
precisely because they couple functionality with executable exploits.

\begin{takeaway}
Evaluation coverage is uneven: benchmarks concentrate on a few CWEs and languages and mostly judge isolated functions, while application-scale, exploit-grounded, functionality-coupled evaluation is recent and scarce.
\end{takeaway}

\newpara{The knowledge blind spot}
What no oracle in this progression measures is whether the model understood the security principle a task demands. This blind spot is not incidental; it leaves four things unmeasured at once. The benchmarks do not ask whether a model \emph{knows} the SCPs behind a target vulnerability; they do not \emph{align} a vulnerability label to the principle needed to avoid it; they do not check whether a model's stated security reasoning is \emph{actuated} in the code it then writes; and they cannot \emph{decompose} a failure into its cause. A single pass/fail verdict collapses sharply different situations---a model that never knew the principle, one that knew it but did not recognize it applied, one that applied the wrong or a partial mitigation, and one that over-secured the code and broke it---into the same red mark. These causes have opposite implications for training, prompting, and benchmark design, yet output-centric measurement renders them indistinguishable.

\begin{takeaway}
Across measurement and interventions alike, the field reasons about secure-coding knowledge everywhere but measures it almost nowhere: evaluation remains output-centric, able to say whether code is secure but not why. This is the deepest pitfall in the field today, and the one the rest of this paper addresses---a principle-centered account that separates what a system \emph{knows} from what it \emph{enacts}.
\end{takeaway}


\section{The \tool{} Framework}
\label{sec:kauge}

\subsection{Design Rationale}
\label{sec:kauge_design}

The survey above shows that secure-code-generation research has advanced across different intervention points. Section~\ref{sec:measurement}, however, exposes a persistent measurement gap. Most evaluation oracles judge the artifact a model emits, rather than the security knowledge and reasoning that the artifact is expected to instantiate. As a result, a single pass/fail verdict can collapse distinct failure modes into the same outcome: the model may lack the relevant principle, fail to recognize that it applies, implement a partial or misplaced mitigation, or overcorrect in a way that blocks the exploit only by breaking required functionality.

\tool{} is designed to make these distinctions measurable. To do so, it needs an evaluation unit that connects three views of secure generation: what security knowledge is available to the model, what code the model produces, and whether the code contains evidence of a plausible defensive mechanism. We use secure coding principles as this unit.

SCPs from established catalogs such as OWASP~\cite{owasp_scp} and CERT~\cite{cert_c} describe intended secure behavior as prescriptive rules: 
they state what developers should do to avoid insecure code, rather than only naming the weakness that appears after a failure~\cite{owasp_scp,realsecbench2026}. This makes them useful for studying language models. Because SCPs are expressed in natural language, they can be posed \textit{as concepts that a model may recognize, reason about, and apply during generation}. At the same time, because they describe coding practices rather than only vulnerability labels, they can be linked to concrete defensive mechanisms in code.


SCPs also provide a more suitable bridge than CWE labels for measuring knowledge-to-code translation. A broad principle, such as input validation, can apply across many CWEs, while a single CWE may require several principles to be applied together~\cite{owasp_scp,cweval}. 
Such relationships are precisely why output labels alone are insufficient: knowing that a task involves command injection, path traversal, or XSS does not by itself identify which practice must be applied, where it must be applied, or whether the implementation preserves functionality. 
In \tool{}, SCPs serve as diagnostic anchors across the generation process: they let us ask whether a model recognizes the relevant practice, whether the generated artifact blocks the exploit while preserving functionality, and whether the code contains evidence of the defensive mechanism that the practice suggests.

\begin{table*}[t]
\centering
\small
\caption{Layer-1 secure-coding knowledge dimensions and their NLP task instantiations. Superscripts mark the source task family: $\mathrm{S}$ = SuperGLUE~\cite{wang2019superglue}, $\mathrm{G}$ = GLUE~\cite{wang2018glue}, $\mathrm{C}$ = CosmosQA~\cite{huang2019cosmosqa}, and $\mathrm{R}$ = ReClor~\cite{yu2020reclor}.}
\label{tab:l1-dimension-defs}
\resizebox{\textwidth}{!}{%
\begin{tabular}{>{\centering\arraybackslash}m{2.0cm} m{3.0cm} m{3.2cm} m{5.4cm} >{\centering\arraybackslash}m{2.0cm}}
\toprule
\textbf{Dimension} & \textbf{Question} & \textbf{Formats} & \textbf{What the formats probe} & \textbf{Metric} \\
\midrule
\cellcolor{SOKBlue!18}\textbf{Declarative} & What a practice states & WiC$^{\mathrm{S}}$; QNLI$^{\mathrm{G}}$, MRPC$^{\mathrm{G}}$ & Term sense, answer entailment, paraphrase equivalence & Accuracy/F1 \\
\cellcolor{SOKRed!18}\textbf{Causal} & Why violations matter & MNLI$^{\mathrm{G}}$, COPA$^{\mathrm{S}}$ & Entailment and cause/effect reasoning & Accuracy \\
\cellcolor{SOKGreen!18}\textbf{Procedural} & How to apply a practice & CosmosQA$^{\mathrm{C}}$, ReClor$^{\mathrm{R}}$ & Commonsense and logical application of a rule & Accuracy \\
\cellcolor{SOKPurple!18}\textbf{Contextual} & When a practice applies & BoolQ$^{\mathrm{S}}$, MultiRC$^{\mathrm{S}}$ & Situation-level relevance and multi-sentence reading & Accuracy/F1 \\
\bottomrule
\end{tabular}
}
\end{table*}

\begin{figure*}[t]
\centering
\subfloat[OWASP]{\includegraphics[width=0.32\textwidth]{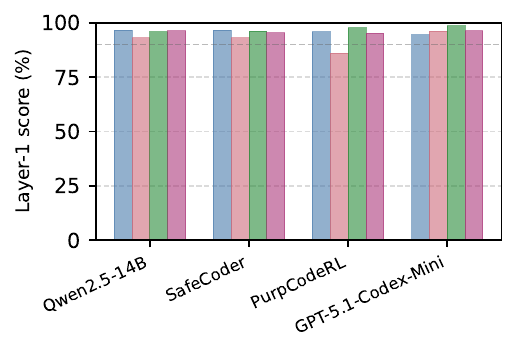}}\hfill
\subfloat[CERT C]{\includegraphics[width=0.32\textwidth]{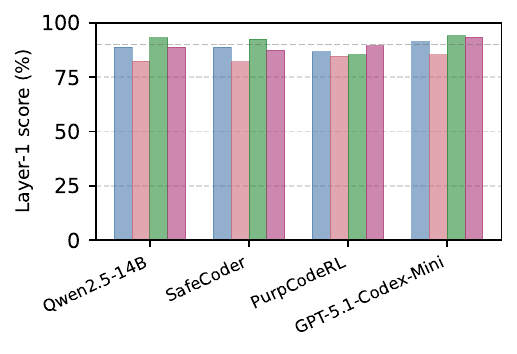}}\hfill
\subfloat[CERT C++]{\includegraphics[width=0.32\textwidth]{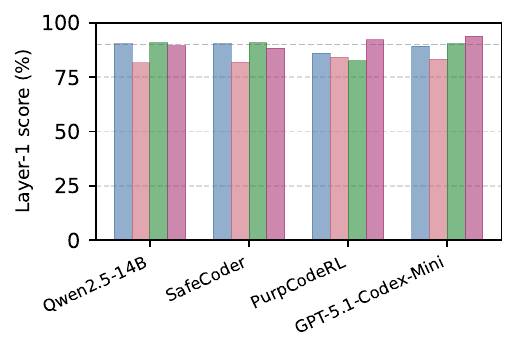}}
\caption{Layer-1 secure-coding knowledge by reasoning dimension, source catalog, and model. Bar colors indicate the four reasoning dimensions: \legendbox{SOKBlue!45}\,Declarative, \legendbox{SOKRed!45}\,Causal, \legendbox{SOKGreen!45}\,Procedural, and \legendbox{SOKPurple!45}\,Contextual.}
\label{fig:l1-dimensions-by-source}
\end{figure*}

\subsection{Framework Overview}
\label{sec:kauge_overview}

The framework is organized around three measurements that give \tool{} its name: \emph{knowledge}, \emph{actuation}, and the \emph{gap} between them. Each layer answers a different question about secure code generation; the methodology for each is described in detail in Sections~\ref{sec:layer1}--\ref{sec:layer3} respectively.

\begin{itemize}

\item \textbf{Knowledge (Layer~1)} measures whether the model can reason about SCPs in natural language. It asks whether the model can identify what a principle requires, explain why violating it creates a vulnerability, determine when the principle is relevant, and describe how it should be applied (Section~\ref{sec:layer1}).

\item \textbf{Actuation (Layer~2)} measures code-level secure generation. It asks whether the model produces programs that satisfy the functional specification and block the corresponding exploit under executable benchmark oracles (Section~\ref{sec:layer2}).

\item \textbf{Gap (Layer~3)} connects the first two layers through mechanism-level diagnosis. It asks whether successful actuation follows the expected defense strategy, and whether failed actuation reflects a missing principle, a partial or misplaced implementation, or security achieved through some other mechanism (Section~\ref{sec:layer3}).

\end{itemize}


Knowledge and Actuation measure two different forms of progress; the Gap explains why they diverge. In this sense, \tool{} is \textit{a diagnostic lens for locating where secure-code generation fails}: in recognizing the relevant practice, translating it into functional code, or implementing the right defense at the right boundary.

\subsection{Evaluation Design}
\label{sec:eval_design}


We choose models and interventions to support diagnosis rather than to construct a single leaderboard. {Qwen2.5-14B-Instruct} serves as the open-weight backbone: it is large enough for nontrivial code generation, small enough to run reproducibly, and shared across the training-time interventions we study. {GPT-5.1-Codex-Mini} provides a proprietary code-specialized reference point.

We study interventions at three levels. At \emph{training time}, {SafeCoder}~\cite{safecoder2024} applies supervised security fine-tuning to the Qwen2.5 backbone, while {PurpCodeRL}~\cite{liu2026purpcode} applies security-oriented reinforcement learning post-training; this trio lets us ask what changes when the backbone is held fixed. At \emph{inference time}, we compare base generation, {Security-Aware prompting}~\cite{tony2025prompting} and {recursive criticism and improvement (RCI)}~\cite{kim2023rci,tony2025prompting}. SCP-guided prompting is treated as a gap intervention: the question it answers is not only whether it changes benchmark outcomes, but whether naming the relevant practice causes the model to implement it. At the \emph{environment level}, the {Codex Scaffold} gives GPT-5.1-Codex-Mini access to a file system, shell, and development environment; it is an actuation intervention because it changes the setting in which the model writes code, not the model's underlying security understanding.

We measure the effect of these interventions on actuation using two executable benchmarks. {CWEval}~\cite{cweval} tests isolated programming tasks with functionality and security tests, close to the function-completion setting. {BaxBench}~\cite{baxbench} tests full web applications across framework environments, pairing functional tests with exploit tests. Together, they provide a function-level view and an application-level view. We use executable oracles rather than static-analysis-based benchmarks because
static analysis is prone to both false positives and false negatives~\cite{10400834,10.1145/3660772,10.1145/3611643.3616262}.


\section{Progress: Secure-Coding Knowledge}
\label{sec:layer1}

\newpara{Goal}
The first empirical question \tool{} asks: can current systems recognize and reason about secure-coding principles in natural language before they are asked to write code?
This matters because many interventions assume that the model can use a security principle once it is named. A zero-shot prompt may say ``validate all inputs.'' An SCP-guided prompt may list a more specific rule. A critique prompt may ask the model to look for injection risks. If the model does not understand what these practices mean, when they apply, or why they matter, then failures in generated code may simply reflect missing knowledge. If the model does understand them, then failures downstream are more revealing: they point to difficulty activating or implementing knowledge rather than merely acquiring it.

\newpara{Diagnostic task}
We operationalize Layer~1 with a natural-language benchmark that measures SCP recognition and reasoning across four reasoning dimensions, summarized in Table~\ref{tab:l1-dimension-defs}: \colorbox{SOKBlue!25}{\emph{declarative}} (what a practice states), \colorbox{SOKRed!25}{\emph{causal}} (why violating it creates a vulnerability), \colorbox{SOKGreen!25}{\emph{procedural}} (how to apply it), and \colorbox{SOKPurple!25}{\emph{contextual}} (whether it is relevant in a concrete situation). These dimensions matter because secure coding knowledge is not just terminology: a model can repeat a rule, yet fail to know why it matters, when it applies, or what implementation step it requires. Each dimension is instantiated using established NLP task formats drawn from GLUE, SuperGLUE, CosmosQA, and ReClor~\cite{wang2018glue,wang2019superglue,williams-etal-2018-broad,huang2019cosmosqa,yu2020reclor}; see Table~\ref{tab:l1-dimension-defs} for the format-to-dimension mapping.

To instantiate this diagnostic task, we crawl and normalize source rules from OWASP and CERT catalogs, retaining the rule identifier, title, source text, language or standard, and domain grouping. For each rule, an LLM generator instantiates fixed-answer questions in NLP task formats matched to the intended reasoning dimension; prompts require each item to be self-contained and scoped to the target practice. Every generated item is flattened into a standard record with a reference answer, task format, rule ID, and reasoning dimension. A separate LLM verifier then checks whether the item is well-formed, self-contained, answerable, and compatible with the stated task format; failed items are omitted from the verified set.

The pipeline produces 6,382 verified fixed-answer questions from 456 source rules: 3,419 from 218 OWASP Secure Coding Practices, 1,967 from 157 CERT C rules, and 996 from 81 CERT C++ rules. Each principle is converted into discrete questions with fixed gold answers. Because the construction is LLM-assisted, we additionally validate a stratified sample of 100 questions for question validity, reference-answer correctness, dimension match, task-format match, and rule alignment (Appendix~\ref{app:human-validation}).

\newpara{Scoring}
Each question uses the standard metric for its NLP task format---accuracy, F1, or MultiRC F1---as defined in the original task papers~\cite{wang2018glue,wang2019superglue,williams-etal-2018-broad,huang2019cosmosqa,yu2020reclor}; Table~\ref{tab:l1-dimension-defs} maps each format to its dimension and metric. Dimension scores are macro-averaged across constituent formats. High Layer~1 scores do not prove that a model will generate secure code, but low scores would make secure generation unlikely.

\newpara{Findings}
Figure~\ref{fig:l1-dimensions-by-source} shows that all four systems score well above chance across all catalogs and dimensions. The lowest mean score in the entire evaluation is 86.2\% (PurpCodeRL on CERT C++), and most systems exceed 88\% even on the harder CERT catalogs. On OWASP, scores cluster between 93--97\%, close to the ceiling for fixed-answer NLP tasks where human performance typically falls in the 95--99\% range. These reasoning scores matter as a baseline because they are not borderline results that could be explained by guessing, format familiarity, or surface pattern matching. Secure-coding knowledge is broadly present before code generation begins.

\begin{takeaway}
Secure-coding knowledge is broadly present across all systems and catalogs. All systems score well above chance, so later code failures cannot be explained simply as absence of the relevant concepts.
\end{takeaway}

\begin{table*}[t]
\centering
\scriptsize
\caption{Layer-2 actuation results showing interventions across models.}
\label{tab:l2-models}
\resizebox{\textwidth}{!}{%
\begin{tabular}{llrrrrrr}
\toprule
\textbf{LLM} & \textbf{Method} & \multicolumn{3}{c}{\textbf{CWEval}} & \multicolumn{3}{c}{\textbf{BaxBench}} \\
\cmidrule(lr){3-5}\cmidrule(lr){6-8}
 & & \textbf{pass@1} & \textbf{secure-pass@1} & \textbf{secure@pass1} & \textbf{pass@1} & \textbf{secure-pass@1} & \textbf{secure@pass1} \\
\midrule
\multirow{5}{*}{Qwen2.5-14B} & Base & \cellcolor{SOKBlue!38}68.9 & \cellcolor{SOKGreen!25}41.2 & \cellcolor{SOKPurple!33}59.8 & \cellcolor{SOKBlue!12}12.8 & \cellcolor{SOKGreen!8}3.8 & \cellcolor{SOKPurple!20}30.0 \\
 & Security-Aware prompt & \cellcolor{SOKBlue!37}68.1 & \cellcolor{SOKGreen!26}42.9 & \cellcolor{SOKPurple!35}63.0 & \cellcolor{SOKBlue!12}12.2 & \cellcolor{SOKGreen!8}5.1 & \cellcolor{SOKPurple!25}41.7 \\
 & RCI & \cellcolor{SOKBlue!28}48.7 & \cellcolor{SOKGreen!23}37.0 & \cellcolor{SOKPurple!41}75.9 & \cellcolor{SOKBlue!12}13.5 & \cellcolor{SOKGreen!9}7.4 & \cellcolor{SOKPurple!31}54.7 \\
 & SafeCoder & \cellcolor{SOKBlue!36}65.5 & \cellcolor{SOKGreen!25}40.3 & \cellcolor{SOKPurple!34}61.5 & \cellcolor{SOKBlue!14}17.1 & \cellcolor{SOKGreen!10}8.4 & \cellcolor{SOKPurple!29}49.3 \\
 & PurpCodeRL & \cellcolor{SOKBlue!30}51.3 & \cellcolor{SOKGreen!20}30.3 & \cellcolor{SOKPurple!33}59.0 & \cellcolor{SOKBlue!11}9.9 & \cellcolor{SOKGreen!8}4.6 & \cellcolor{SOKPurple!27}46.2 \\
\midrule
\multirow{4}{*}{GPT-5.1-Codex-Mini} & Base & \cellcolor{SOKBlue!45}84.9 & \cellcolor{SOKGreen!34}60.5 & \cellcolor{SOKPurple!39}71.3 & \cellcolor{SOKBlue!30}52.6 & \cellcolor{SOKGreen!22}34.2 & \cellcolor{SOKPurple!36}65.0 \\
 & Security-Aware prompt & \cellcolor{SOKBlue!44}82.4 & \cellcolor{SOKGreen!39}72.3 & \cellcolor{SOKPurple!46}87.8 & \cellcolor{SOKBlue!31}53.8 & \cellcolor{SOKGreen!24}38.3 & \cellcolor{SOKPurple!39}71.1 \\
 & RCI & \cellcolor{SOKBlue!33}59.7 & \cellcolor{SOKGreen!32}57.1 & \cellcolor{SOKPurple!50}\textbf{95.8} & \cellcolor{SOKBlue!24}38.5 & \cellcolor{SOKGreen!20}30.6 & \cellcolor{SOKPurple!43}\textbf{79.5} \\
 & Codex Scaffold & \cellcolor{SOKBlue!50}\textbf{96.6} & \cellcolor{SOKGreen!48}\textbf{91.6} & \cellcolor{SOKPurple!50}94.8 & \cellcolor{SOKBlue!33}\textbf{58.2} & \cellcolor{SOKGreen!25}\textbf{41.8} & \cellcolor{SOKPurple!39}71.9 \\
\bottomrule
\end{tabular}%
}
\end{table*}

The more informative variation is across catalogs and dimensions rather than across models. OWASP mean scores (93.7--96.5\%) sit roughly 7--9 points above CERT C (86.6--91.1\%) and CERT C++ (86.2--89.2\%), consistent with OWASP's higher-level web-oriented rules being closer to general training distribution than low-level C/C++ memory and concurrency rules. Within the CERT catalogs, causal reasoning is the clearest stress point: across all four models and both CERT catalogs, the causal dimension score is the lowest of the four dimensions in every case, a pattern that holds without exception (p=0.008, sign test). Models can often recognize what a secure coding rule says and often know when it applies, but the \emph{why} behind language-specific low-level rules remains harder. This matters downstream because many code-generation failures are not failures to name a rule; they are failures to preserve the vulnerability rationale while satisfying an implementation contract.

Security fine-tuning does not substantially change this picture. SafeCoder tracks the Qwen2.5 base within 1 point across all catalogs, and PurpCodeRL is slightly weaker on OWASP (93.7\% vs.\ 95.5\%) despite being security-oriented. This does not mean security tuning is ineffective for code generation; rather, it suggests that these interventions do not primarily improve the measured natural-language recognition of SCPs. Their effects, if any, should therefore appear more clearly in code-level actuation.

\begin{takeaway}
Security knowledge gaps are driven by reasoning type and rule domain, not
security specialization. Models handle high-level web-oriented principles
well but struggle with causal reasoning over low-level C/C++ rules.
Security fine-tuning does not close this gap.
\end{takeaway}


\section{Progress: Code-Level Actuation}
\label{sec:layer2}

\newpara{Goal}
Having established that secure coding concepts are broadly understood by current models, Layer~2 asks whether that understanding translates into secure code in practice.

\newpara{Diagnostic task}
We measure actuation on two executable benchmarks. {CWEval}~\cite{cweval} tests isolated programming tasks with functionality and security tests, covering 119 tasks across 5 languages (C, C++, Go, JavaScript, Python) and 31 CWEs. {BaxBench}~\cite{baxbench} tests full web applications across framework environments, pairing functional tests with exploit tests; it comprises 392 scenario-framework tasks spanning 28 scenarios, 14 environments, 6 languages (Go, JavaScript, PHP, Python, Ruby, Rust), and 13 CWEs, with a mix of single-file and multi-file projects. Together they provide a function-level view and an application-level view of secure generation. For each model and method, we generate one sample per task with the default benchmark settings.

\newpara{Scoring}
We adopt the metrics introduced by CodeGuard+~\cite{fu2024constrained}. Given a benchmark of $N$ tasks, we generate $n \geq k$ samples per task and count, for each task, the number of samples $c$ that pass the functional oracle and the number $sp$ that pass both the functional and security oracles.

The three metrics are:
\begin{align}
  \passatk{} &:= \mathbb{E}_{\text{tasks}}\!\left[1 - \frac{\binom{n-c}{k}}{\binom{n}{k}}\right], \label{eq:pass-at-k}\\[4pt]
  \securepassatk{} &:= \mathbb{E}_{\text{tasks}}\!\left[1 - \frac{\binom{n-sp}{k}}{\binom{n}{k}}\right], \label{eq:secure-pass-at-k}\\[4pt]
  \seccondk{} &:= \mathbb{E}_{\text{tasks}}\!\left[1 - \frac{\binom{c - sp}{k}}{\binom{c}{k}}\right], \label{eq:secure-at-pass-k}
\end{align}
where \passatk{}, \securepassatk{}, and \seccondk{} measure functionality, joint functional-security correctness, and security conditional on functionality, respectively.


\newpara{Findings}
Table~\ref{tab:l2-models} reports all three metrics for base systems and actuation interventions. A natural first question is whether Layer~1 knowledge aligns with Layer~2 actuation. Four systems are too few to support a strong correlation claim, so we expand this analysis to 10 base models with Layer~1 measurements in Appendix~\ref{app:ka-base}. Across these 10 models, higher Layer~1 knowledge is associated with higher downstream \securepassatk{}. The Spearman rank correlation is significant for CWEval \securepassatk{} \stat{($n{=}10$, $\rho{=}0.71$, $p{=}0.025$)} and directionally consistent for BaxBench \securepassatk{} \stat{($n{=}10$, $\rho{=}0.63$, $p{=}0.056$)}. Knowledge is therefore a meaningful predictor of actuation, particularly at function level, but it is not sufficient to explain whole-application behavior, where project construction quality also plays an important role.

\smallskip\noindent\textit{Base models.}
Layer~1 knowledge is high, but actuation is much harder. On CWEval, the Qwen2.5 base run splits into three comparably important buckets: functional-and-secure solutions, functional-but-insecure solutions, and nonfunctional solutions. GPT-5.1-Codex-Mini reduces nonfunctional failures, but functional code can still remain exploitable, especially for familiar vulnerability classes such as path traversal (CWE-22), XSS (CWE-79), SSRF (CWE-918), and log injection (CWE-117).

At application scale, the same problem becomes even sharper: many generations fail before security can be meaningfully tested because they do not produce a runnable backend satisfying the API contract. Here, functionality is a prerequisite for observing security at all. This is why we avoid standalone security rates: treating applications with no recorded CWE as secure would over-credit broken systems whose exploit tests simply could not run. Among functional-but-insecure applications, failures again concentrate in familiar classes, including resource exhaustion (CWE-400), path traversal (CWE-22), log injection (CWE-117), code injection (CWE-94), missing authorization (CWE-284), XSS (CWE-79), and insecure credential handling (CWE-522).

\begin{takeaway}
Actuation failures span both functionality and security: models may generate code that does not run, code that meets the functional contract but remains exploitable, or deployable code that satisfies both. Secure-code evaluation should therefore report joint functional and security correctness rather than crediting broken programs as secure or treating functional correctness alone as sufficient.
\end{takeaway}

\smallskip\noindent\textit{Inference-time interventions.}
Security-Aware prompting helps GPT-5.1-Codex-Mini most on CWEval. In a paired comparison against the base prompt, 20 tasks switch from failing to passing the CWE security test, while 6 switch from passing to failing \stat{($p{=}0.009$)}. These gains come primarily from tasks that were already functional but insecure under the base prompt, suggesting that the prompt helps the model apply defenses in code that was otherwise close to correct. The corresponding paired functionality comparison is not significant: 6 tasks become functional, while 9 previously functional tasks become nonfunctional \stat{($p{=}0.607$)}.

RCI illustrates a more aggressive inference-time tradeoff. It achieves the highest \seccondk{} scores of any inference-time method, but often by breaking previously functional code for both models \stat{($p{<}.001$)}. The joint functional-and-secure change is not significant for either model. The code-level examples illustrate why. In a CWE-78 command-injection task, the Qwen2.5 RCI solution avoids the shell by replacing the required \texttt{ls -l} behavior with a filesystem directory iterator, which blocks the injection pattern but fails the functionality test. In a CWE-329 predictable-IV task, GPT-5.1-Codex-Mini's RCI solution upgrades from AES-CBC to AES-GCM, a security-conscious change that breaks the benchmark's CBC-based decryption contract. Both are instances of over-actuation: the model applies a plausible security improvement without preserving the requested behavior. These changes improve the security oracle while moving the implementation away from the specified task.

\begin{takeaway}
Inference-time interventions act as steering mechanisms rather than new capabilities: they can recover security behavior when the base model is close to correct, but they can also over-steer and disturb the requested program behavior. Their benefit is therefore local and task-dependent, not a reliable substitute for stronger actuation.
\end{takeaway}

\smallskip\noindent\textit{Training-time interventions.}
The fine-tuning rows are similarly mixed. SafeCoder helps most on BaxBench, where improvements in framework-correct implementation have room to matter, but does not clearly improve function-level performance on CWEval. PurpCodeRL underperforms the base model on joint correctness in both settings. Security specialization, therefore, does not automatically transfer into robust executable behavior across benchmark granularities: a security-trained model can still lose if the training intervention weakens general implementation competence or overfits to narrower security idioms than the benchmark requires.


\begin{takeaway} Security training improves actuation only when it preserves general implementation competence; otherwise, specialization can weaken functionality or overfit to narrow security idioms. \end{takeaway}

\smallskip\noindent\textit{Agentic scaffolding.}
The Codex Scaffold is the strongest on both benchmarks. The result indicates that the gain does not reflect additional secure-coding knowledge but evidence that file-system access, shell use, and project-level generation improve actuation. The gain is largest where implementation completeness matters most, supporting the claim that secure code generation is partly a software-construction problem, not only a security-knowledge problem.

\begin{takeaway}
Agentic scaffolding improves actuation by strengthening software construction. Better project-level generation can carry security along with functionality, even without changing the model's underlying security knowledge.
\end{takeaway}

\begin{figure*}[t]
\centering
\includegraphics[width=\textwidth]{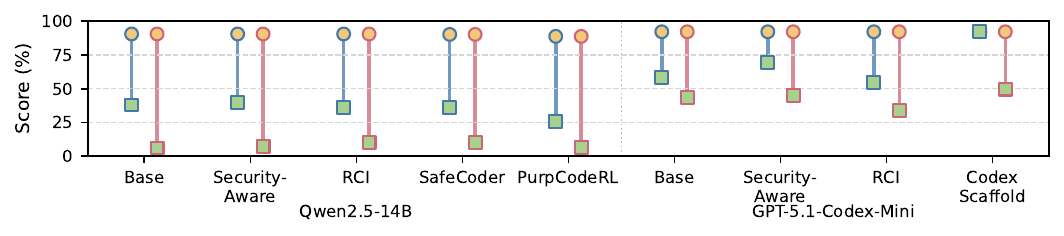}
\vspace{-6mm}
\caption{Knowledge--actuation gap by benchmark and intervention, using the same model/method slices as Table~\ref{tab:l2-models}. $K$ is the Layer-1 secure-coding knowledge score of the underlying model; $A$ is \textsf{secure-pass@1}, the joint rate at which generated code is both functional and secure. \textcolor{SOKGold}{$\bullet$}\,$K$ and \textcolor{SOKLightGreen}{$\blacksquare$}\,$A$ are connected by benchmark: \legendline{SOKBlue}\,CWEval, \legendline{SOKRed}\,BaxBench.}
\label{fig:l3-ka-gap}
\end{figure*}

\section{Pitfalls: The Knowledge--Actuation Gap}
\label{sec:layer3}

\newpara{Goal}
Layer~1 and Layer~2 establish that models broadly understand secure coding practices and that many interventions improve the rate of functional, secure code generation. The pitfall is that these are not the same accomplishment. A model can recognize a principle in natural language yet fail to carry it into code; it can also produce secure code for a particular task without that success implying reliable principle actuation. Layer~3 measures the distance between the two, using SCP compliance to diagnose what kind of actuation actually occurred and where it broke down.

\newpara{Diagnostic task}
To diagnose the gap, we compare SCP compliance against the executable benchmark verdict using the same CWEval and BaxBench benchmarks as Layer~2. For each exploit, our mapping pipeline identifies the relevant secure coding practices and groups them into one or more sufficient defense strategies, represented as a Disjunctive Normal Form (DNF) formula: a generated solution is SCP-compliant if the judge finds that it implements every SCP in at least one sufficient clause. The map contains 176 exploit-level formulas (108 for CWEval, 68 for BaxBench), covering 1--10 relevant SCPs per exploit with a mean of 3.0. Crossing SCP compliance with benchmark security produces four diagnostic outcomes for functional code: 
\colorbox{SOKLightBlue!60}{\emph{principled success}} (secure and compliant), \colorbox{SOKLightGreen!60}{\emph{secure by other means}} (secure but not compliant), \colorbox{SOKGold!60}{\emph{compliant but vulnerable}} (compliant but exploitable), and \colorbox{SOKPink!60}{\emph{actuation failure}} (exploitable and not compliant). \colorbox{SOKLightGray!60}{Nonfunctional code} is tracked separately.

To build the exploit-to-SCP map, two independent judges, {GPT-5-nano} and {Claude Haiku 4.5} (\texttt{claude-haiku-4-5-20251001}), receive the task, concrete exploit, CWE, and candidate SCPs, and produce minimal alternative defense strategies. Each strategy is an AND clause and the full formula is an OR over distinct sufficient strategies. We use two different model providers to reduce single-model leniency, with a {GPT-5-nano} adjudicator reconciling disagreements into the final DNF. To judge generated code, the same two-judge setup checks whether the code implements each relevant SCP, again with GPT-5-nano adjudicating per-SCP disagreements. CWEval is judged as one code artifact per exploit; BaxBench is judged per exploit over cleaned project files, with file-level SCP evidence unioned before evaluating the DNF. We judge SCP compliance only for functional generations, because broken programs are not meaningful evidence of exploit-level principle actuation. Because both steps are LLM-assisted, we validate stratified samples of 100 items each for the exploit-to-SCP formula mapping and the SCP-compliance judge; results are reported in Appendix~\ref{app:human-validation}.

\newpara{Scoring}
We define two summary scores. \emph{Knowledge} ($K$) is the macro-averaged Layer~1 score of the underlying model across all reasoning dimensions and source catalogs, capturing how well the model understands secure coding practices as natural-language concepts. \emph{Actuation} ($A$) is \securepassatk{} from Layer~2: the joint rate at which generated code is both functional and secure.


\newpara{Findings}
Figure~\ref{fig:l3-ka-gap} makes the gap $K{-}A$ visible. Knowledge and actuation move in the same direction across slices (Spearman rank correlation: \stat{$n=18$, $\rho=0.75$, $p{<}.001$}), but actuation is systematically lower than knowledge across every slice (Wilcoxon signed-rank test: \stat{$n=18$, $W^+=170$, $p{<}.001$}). The benchmark-level gap is 47.9 points for CWEval and 72.9 points for BaxBench. CWEval isolates small function-level tasks, so a single missing guard can directly separate knowing from doing; BaxBench asks for project-level web applications, where actuation depends on framework defaults, routing, middleware, and file organization as well as the local security concept. Application-scale actuation mixes security knowledge with software-construction quality, which is why the gap cannot be closed by adding more security vocabulary alone.

\begin{takeaway}
The central pitfall is the knowledge--actuation gap. The models usually know the relevant security concepts, but that knowledge drops substantially when it must become functional, exploit-resistant code.
\end{takeaway}

\begin{figure*}[t]
\centering
\includegraphics[width=\textwidth]{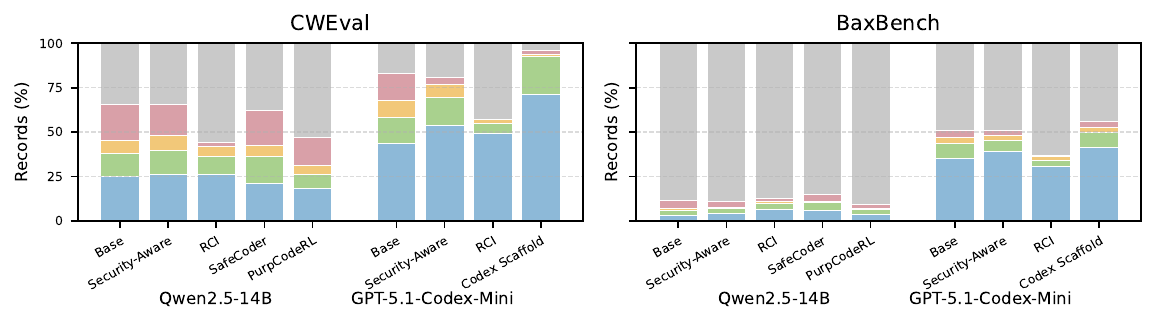}
\vspace{-6mm}
\caption{Layer-3 outcome distribution by system, method, and benchmark, using the same model/method slices as Table~\ref{tab:l2-models}. Each vertical stacked bar covers all exploit-level records for a slice. From bottom to top: \legendbox{SOKLightBlue}\,principled success, \legendbox{SOKLightGreen}\,secure by other means, \legendbox{SOKGold}\,compliant but vulnerable, \legendbox{SOKPink}\,actuation failure, and \legendbox{SOKLightGray}\,nonfunctional code.}
\label{fig:l3-taxonomy-bars}
\end{figure*}

Figure~\ref{fig:l3-taxonomy-bars} shows how the diagnostic outcomes distribute across systems and methods. Each case reveals a different relationship between knowledge, compliance, and security --- and three of the four examples below come from \texttt{qwen2.5-safecoder}, a model fine-tuned for security, which makes the point concrete: security tuning does not close the actuation gap, it shifts where failures land.

\colorbox{SOKLightBlue!60}{\emph{Principled success}} is the desired outcome and the majority case: the model knows the relevant practice, implements it correctly, and the exploit fails. In a CWE-943 query-injection task, \texttt{qwen2.5-safecoder} applies the mapped parameterized-query practice correctly, binding username and password as separate parameters rather than interpolating them into the query string. Knowledge, compliance, and security align. This case dominates for vulnerability classes with a single, well-defined defense --- SQL and query injection chief among them.

\colorbox{SOKLightGreen!60}{\emph{Secure by other means}} arises when code resists an exploit through a mechanism outside the mapped defense strategy. In a CWE-22 path-traversal task, the Codex agent implements a containment check that verifies the resolved target path stays within the destination directory, rather than the mapped allow-list or explicit canonicalization pattern. The exploit fails, but the defense lies outside the mapped formula, so the judge scores it non-compliant. This is not a model failure but a coverage limit: the defense exists, just not where the SCP formula expects it. In BaxBench this cell is driven by the same dynamic at larger scale --- framework-level path normalization, authorization middleware, and project-level defaults block exploits with no compact sink-level pattern to match.

\colorbox{SOKGold!60}{\emph{Compliant but vulnerable}} is the most instructive failure: the model recognizes and attempts the right practice, but the abstract SCP underdetermines its secure implementation. In a CWE-113 HTTP response-splitting task, \texttt{Safecoder} allow-lists the header type and validates the user-controlled value --- but checks whether it is UTF-8 encodable rather than whether it contains CR/LF control characters. Since \texttt{\textbackslash r}/\texttt{\textbackslash n} are perfectly valid UTF-8, the CRLF injection passes straight through. The model validated the wrong property and missed the one-line fix of rejecting or stripping control characters. The judge credits it as compliant with the input-validation and output-encoding principles, yet the code remains exploitable. The principle is too abstract to guarantee the sink-specific fix.

\colorbox{SOKPink!60}{\emph{Actuation failure}} is the cleanest case: functional, exploitable, and no relevant practice applied at all. In a CWE-79 XSS task, the same \texttt{Safecoder} base run interpolates user-controlled input directly into an HTML-rendered string with no escaping. This is not a knowledge gap --- the model can identify and explain output encoding --- but a failure to recognize that it applies here.

The imbalance between secure-only and compliant-only cases is statistically sharp: \stat{secure-only 1{,}100 vs.\ compliant-only 287, $p{<}.001$} across all functional exploit-level records, confirming that security and compliance diverge in systematic rather than random ways.

\begin{takeaway}
SCP compliance does not define the knowledge--actuation gap; it diagnoses it. The same actuation score can hide \colorbox{SOKLightBlue!60}{\emph{principled success}}, \colorbox{SOKLightGreen!60}{\emph{secure-by-other-means}} behavior where the defense lives outside the mapped formula, \colorbox{SOKGold!60}{\emph{compliant but vulnerable}} code where an abstract principle underdetermines its sink-specific implementation, and \colorbox{SOKPink!60}{\emph{actuation failures}} where the model never applies the relevant practice at all.
\end{takeaway}



Figure~\ref{fig:l3-taxonomy-bars} reveals a more specific pattern than ``more intervention helps.'' The successful methods do not primarily raise the knowledge point; instead, they change how security constraints are carried through the generation process. Lightweight prompting helps when the model is already near the right implementation, recovering security behavior with minimal disruption to functionality. Stronger gains appear when the intervention supplies structure the model otherwise has to invent: preserving the functional contract, locating the relevant boundary, assembling the project, or testing whether exploit-facing behavior actually changed. The Codex Scaffold makes this visible: it does not make GPT-5.1-Codex-Mini more knowledgeable by our Layer~1 measure, but it narrows the K--A distance by giving the model a construction environment in which functionality, integration, and security constraints can be coordinated. In this sense, the gap is not just between knowing and coding; it is about whether the generation process keeps a known principle attached to the right code region, API boundary, framework mechanism, and executable behavior.

\begin{takeaway}
The knowledge–actuation gap is a delivery problem: models hold the right security knowledge but fail to apply it at the correct implementation boundary. Interventions help not by adding more knowledge, but by binding known principles to the right code locations.
\end{takeaway}

\section{Paths Forward}
\label{sec:paths}

The survey in \S\ref{sec:systematization} shows that the field has already explored many ways to move security knowledge closer to generation: adding it to the prompt, internalizing it in model weights, constraining decoding, and feeding back oracle results after a first draft. Our measurements explain why the remaining problem is not solved by any one of these alone. Layer~1 says the concepts are usually available; Layer~2 says secure working code is still much harder than secure-coding recognition; Layer~3 says the missing piece is often the implementation boundary where a principle must become a concrete defense. 
This leads to five concrete directions.

\subsection{Executable Feedback over Security Intent}
The strongest Layer~2 evidence is that scaffolding helps by improving software construction. The Codex Scaffold does not make GPT-5.1-Codex-Mini more knowledgeable in Layer~1; it gives the model project context, file-system access, a shell, and an execution environment for producing runnable artifacts. This matters because many failures are not subtle missing principles, but failures to build working applications before security is even testable. Secure generation should therefore treat execution-grounded development as a first-class setting, not merely a wrapper around a model.


Post-generation feedback points in this direction, but the feedback must be security-specific and executable. In our experiments, agents generated code and were evaluated only afterward; they did not receive exploit failures, SCP-compliance failures, or property-test failures as repair signals. Future systems should close this loop: generate code, run functional and exploit tests, localize the failing sink or boundary, and revise while preserving the functional contract. This extends RL- and oracle-guided work~\cite{xu2024prosec,liu2026purpcode,nunez2024autosafecoder,saul2025scgagent}: the repair or reward signal should optimize the joint outcome, rather than security alone or refusal behavior. Scaling this loop to applications is difficult because exploit tests are expensive, stateful, and framework-dependent, but that is precisely where the gap is largest.

\subsection{Sink-Grounded Principles}

Layer~3 shows why generic security reminders saturate quickly. An SCP such as input validation, output encoding, or safe file handling is too abstract to determine the implementation. The Layer~3 diagnostic categories expose two different ways that compliance and executable security diverge: \colorbox{SOKLightGreen!60}{\emph{secure-by-other-means}} cases show that executable security can arise outside our mapped defense pattern, while \colorbox{SOKGold!60}{\emph{compliant-but-vulnerable}} cases show that a recognizable defense can still miss the exploit.

This points to a different use of SCPs. Instead of presenting principles as prose checklists, systems should compile them into sink-aware implementation guidance: parameterized database queries for SQL construction, canonicalize-then-confine patterns for path traversal, explicit header APIs for response construction, bounded algorithms for resource exhaustion, and framework-specific authorization middleware for access-control checks. Retrieval can help, but only if it retrieves idioms at the same granularity as the exploit. Constrained decoding can help, but only if the constraint represents the relevant API/dataflow boundary rather than a surface pattern. Fine-tuning can help, but only if training examples preserve functionality while showing the exact secure idiom. In short, the principle must arrive as an implementation plan, not merely as a rule name.

\subsection{Functionality-Preserving Principle Injection}
\begin{figure*}[t]
\centering
\includegraphics[width=\textwidth]{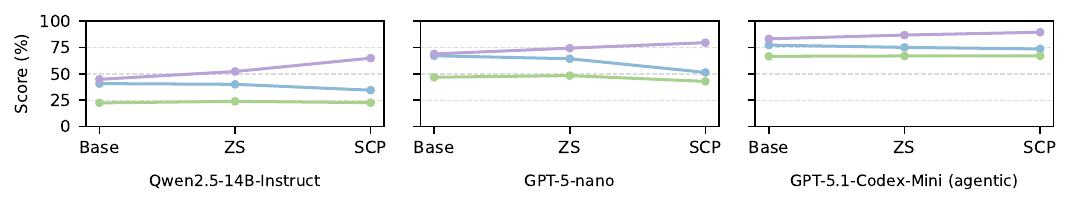}
\vspace{-7mm}
\caption{SCP-guided prompting as a path-forward intervention. Each panel tracks Base $\rightarrow$ zero-shot security prompting (ZS) $\rightarrow$ task-relevant SCP prompting for one system, using a benchmark-balanced aggregate over CWEval and BaxBench.  \legendline{SOKLightBlue}\,\passatk{}, \legendline{SOKLightGreen}\,\securepassatk{}, and \legendline{SOKLightPurple}\,\seccondk{}.}
\label{fig:paths-scp-trend}
\end{figure*}

SCP-guided prompting is a clean way to test whether explicit principles can steer generation, but it also exposes a key risk. Figure~\ref{fig:paths-scp-trend} shows a benchmark-balanced trend from a functionality-oriented base prompt, to a generic Security-Aware prompt, to a task-relevant SCP prompt. The SCP prompt improves \seccondk{} by +12.4 pp relative to Base, meaning that code that remains functional is more often secure. Yet \securepassatk{} changes by only -1.1 pp because functionality can fall at the same time. As with RCI in Layer~2, a security-conscious change can be directionally correct while violating the functional contract.

The implication is not to remove principles from prompts, but to make principle injection functionality-preserving. An SCP-guided system should identify the minimal code region that must change, apply the secure idiom, and preserve the externally visible behavior. For function-level tasks, this means targeted edits around the sink rather than wholesale algorithm replacement. For application-level tasks, it means modifying the relevant route handler, serializer, ORM query, or middleware without disturbing the project scaffold. In this view, prompting names the defense, while tests and localization prevent overcorrection.

\subsection{Mechanism-Grounded Secure-Code Training}
Layer~1 identifies the kind of knowledge that training should strengthen. The most consistent weakness is causal reasoning over lower-level CERT rules: models can often identify a rule, but they are less reliable at explaining why violating it creates the vulnerability. This matters because correct mitigation depends on understanding the source-to-sink vulnerability mechanism. A model that knows ``validate input'' but not why a particular parser, path join, shell invocation, or cipher mode is dangerous may add the wrong validation or change the wrong part of the code.

Future training can target causal actuation: examples and rewards that connect a vulnerability mechanism to the smallest functionality-preserving secure edit. This is different from ordinary supervised secure-code fine-tuning, which can teach a model that secure-looking code is preferred, and different from general RL for vulnerability avoidance, which can reward any artifact that avoids a checker. The desired training signal should say: this attacker input reaches this sink; this SCP applies because it breaks that path; this edit blocks the exploit while preserving the functional tests. That kind of data is more expensive than rule text, but our results suggest it is closer to the missing capability.

\subsection{Principle- and Exploit-Aware Evaluation}
Measurement practice should move beyond single-axis scores, because each metric hides a different failure mode. \seccondk{} can hide functionality collapse; \securepassatk{} can hide whether security came from the mapped defense; SCP compliance can hide whether the defense actually blocks the exploit; and static-analysis findings can scale but reintroduce false positives and false negatives. A mature benchmark should therefore report a compact bundle: \passatk{} for construction, \securepassatk{} for deployable security, \seccondk{} for security conditional on working code, and SCP compliance as a mechanism-level diagnostic view.


The evaluation unit must also match the generated artifact. Our agent results forced this point: whole-project agents cannot be judged by extracting one file. Multi-file applications need a whole-project collection, dependency/cache filtering, and a judgment that can inspect all files relevant to an exploit. This is not a bookkeeping detail; it changes the scientific claim. A single-file judge can miss defenses implemented in routing, middleware, configuration, or helper code, and can over-credit vulnerabilities created by cross-file interactions. Future benchmarks should therefore connect exploits to project-level artifacts while preserving function-level benchmarks for controlled diagnosis.


\section{Threats to Validity}
\label{sec:threats}

\noindent \textbf{Construct and oracle validity}
KAUGE operationalizes secure code generation through constructed artifacts, so each layer captures only part of the broader phenomenon. Layer~1 measures fixed-answer natural-language understanding of SCPs; Layer~2 relies on executable benchmark tests that may miss untested vulnerabilities or encode benchmark-specific assumptions; and Layer~3 uses exploit-to-SCP formulas with an LLM-based compliance judge, whose verdicts are diagnostic rather than ground truth. A formula may omit a valid alternative defense or over-credit a partial one. We mitigate these risks by aligning principles, CWEs, exploits, generated code, and oracles, and by validating questions, formulas, and judge verdicts with stratified human samples.

\noindent \textbf{Scope and generalization}
Our results may not generalize to all models, languages, vulnerability classes, or development settings. We evaluate a bounded set of systems on CWEval and BaxBench, with appendix checks for additional base models, but the study is not intended as a comprehensive leaderboard. These benchmarks cover important exploit patterns, yet real projects involve larger dependency graphs, organization-specific APIs, deployment constraints, and evolving threat models.

\noindent \textbf{Reproducibility and contamination}
Reproducibility is limited by model and benchmark dynamics. Commercial APIs can change, and some models may have seen similar secure-coding examples or benchmark tasks during training. We record model names, access dates, and generation parameters, and rely on executable exploit outcomes where possible, but these steps cannot fully eliminate version instability or contamination.

\section{Related Work}
\label{sec:related}

Recent work~\cite{dai2025rethinking} shows that secure-code-generation studies often evaluate security and functionality separately and rely on limited static-analysis signals, which can over-credit methods that degrade functionality, remove required behavior, or generate task-irrelevant code. Our Layer~2 builds on this critique by treating secure generation as a joint functional-security outcome. KAUGE is complementary: it asks not only whether code is functional and secure, but \textit{why}---whether it implements the mapped secure-coding practice, blocks the exploit through another mechanism, or merely appears compliant while remaining vulnerable.
 
Other related SoKs and surveys have primarily organized LLMs for software security by task, including vulnerability detection, vulnerability repair, and broader AI4Code security risks. Zhou et al.~\cite{zhou2024llmvulrepair} survey LLM-based vulnerability detection and repair, while automated vulnerability repair SoKs systematize vulnerability analysis, patch generation, patch validation, datasets, and tool assessments~\cite{li2025sokavr,hu2025sokavr}. Broader AI4Code security SoKs further discuss risks across code generation, vulnerability detection, and code translation, including insecure outputs, benchmark limitations, data leakage, and robustness concerns~\cite{wu2025sokai4code}. These works are complementary to ours, but they do not center the mechanism-level question studied here: how security knowledge is supplied, internalized, constrained, or fed back, and whether that knowledge is actuated into the right code behavior. This paper therefore complements prior surveys and benchmarks by organizing secure generation around the translation of secure-coding principles into functional, exploit-resistant implementations.

\section{Conclusion}
\label{sec:conclusion}


In this SoK, we systematized AI secure code generation as progress, pitfalls, and paths forward. We surveyed how methods carry security knowledge into generation, measured what models know in Layer~1, evaluated what they enact in Layer~2, and used Layer~3 to diagnose why knowledge still fails to become the right defense at the right implementation boundary. The path forward is to optimize that translation from principle understanding to secure programming.

\bibliographystyle{IEEEtran}
\bibliography{refs}

\balance

\balance

\appendices

\section{Layer-1 Question Formats: Examples and Purpose}
\label{app:l1-format-examples}

Table~\ref{tab:l1-format-examples} gives, for each of the nine NLP task formats, the
cognitive dimension it serves, what it is designed to probe, and an abbreviated example
item. All examples are drawn from OWASP access-control principles for consistency;
the gold answer is shown in brackets.

\begin{table*}[pb!]
\centering
\caption{Layer-1 question formats, the dimension each serves, what it probes, and an
  abbreviated example item.}
\label{tab:l1-format-examples}
\scriptsize
\begin{tabularx}{\textwidth}{@{}l l p{3.4cm} X@{}}
\toprule
\textbf{Format} & \textbf{Dimension} & \textbf{Probes} & \textbf{Abbreviated example (gold)} \\
\midrule
WiC & Declarative & whether a security term carries the same sense across two contexts &
  Does \emph{``enforce authorization on every request''} mean the same in an API-gateway
  context and a nightly-batch context? [true] \\
\addlinespace
QNLI & Declarative & whether a statement answers a principle-identification question &
  Q: which practice restricts protected URLs to authorized users? Sentence: \emph{``the
  app verifies each request against the user's permissions and denies protected URLs
  otherwise.''} [entailment] \\
\addlinespace
MRPC & Declarative & whether two phrasings of a principle are equivalent &
  S1: \emph{``use one site-wide component for all access-authorization checks.''}
  S2: \emph{``centralize all access-control decisions in a single component.''}
  [equivalent] \\
\addlinespace
MNLI & Causal & whether a consequence follows from a principle or its violation &
  Premise: \emph{``deny all access if the app cannot read its security configuration.''}
  Hypothesis: \emph{``if the config cannot be read, every request is blocked.''}
  [entailment] \\
\addlinespace
COPA & Causal & selecting the cause or effect of an insecure behavior &
  An endpoint returns another user's profile without a permission check. Cause?
  (1)~input not sanitized; (2)~\emph{missing authorization check}. [2] \\
\addlinespace
CosmosQA & Procedural & applying a principle in a described scenario &
  Services each implement ad-hoc authorization and some skip checks; best approach?
  (1)~\emph{centralize via one shared authorization component}; \ldots\ [1] \\
\addlinespace
ReClor & Procedural & multi-step logical reasoning over procedural rules &
  Given rules requiring centralized authorization and routing all libraries through it,
  which conclusion follows? [1] \\
\addlinespace
BoolQ & Contextual & yes/no judgment of whether a principle is satisfied in a situation &
  Passage: \emph{processAuditReport} is restricted to auditors and the system checks the
  user's role before invoking it. Is access control applied? [yes] \\
\addlinespace
MultiRC & Contextual & selecting all statements that hold about a situation &
  Passage: a portal and an admin dashboard; a server-side script uses a hard-coded system
  token\ldots\ which statements about the access-control issues are true? [select all] \\
\bottomrule
\end{tabularx}
\end{table*}

\section{Human Validation Protocol}
\label{app:human-validation}

We validate the three constructed artifacts with independent stratified human samples of
$n{=}100$ each. Rates are reported with Wilson 95\% confidence intervals.

\smallskip\noindent\textbf{Layer-1 Question Validation.}
The sample is stratified by standard (CERT~C, CERT~C++, OWASP), reasoning
dimension (declarative, causal, procedural, contextual), and NLP task format.
For each question the annotator checked whether the question is well-formed and
answerable, whether the reference answer is correct, and whether the question
tests the labelled source rule rather than an adjacent practice.
Question validity is 98.0\% (93.0--99.4), reference-answer correctness
93.0\% (86.3--96.6), and rule alignment 84.0\% (75.6--89.9).
Reference-answer correctness is the most load-bearing check; rule alignment
rises to 96.0\% when questions that blend the target rule with an adjacent
practice are counted as partial credit.

\smallskip\noindent\textbf{Exploit--SCP DNF Mapping Validation.}
The sample is stratified by benchmark, DNF resolvability, and CWE bucket.
Each exploit is mapped to a defense formula in Disjunctive Normal Form (DNF):
an OR of AND-clauses, where each clause is a complete defense strategy ---
a minimal set of SCPs that together are sufficient to block the exploit.
The annotator checked two properties per mapping. First, whether at least one
clause describes a genuinely sufficient defense, meaning that a generated
solution implementing all SCPs in that clause would block the exploit while
preserving functionality --- this is the key soundness property for the
judge's ground truth. Second, whether any necessary SCP is absent from the
formula entirely. A sufficient defense clause was confirmed on 87.0\%
(79.0--92.2) of mappings. A necessary SCP was judged missing on only
12.0\% (7.0--19.8), bounding the false-negative rate of the ground-truth
formulas.

\smallskip\noindent\textbf{SCP-Compliance Judge Validation.}
The sample is balanced across the four functional-outcome strata in Section~\ref{sec:layer3} --- \colorbox{SOKLightBlue!60}{\emph{principled success}},
\colorbox{SOKLightGreen!60}{\emph{secure by other means}},
\colorbox{SOKGold!60}{\emph{compliant but vulnerable}}, and
\colorbox{SOKPink!60}{\emph{actuation failure}} --- with 25 items drawn from
each. These strata cover all functionally-correct generations, i.e.\ code that
runs and meets the specification, so that judge behavior is evaluated on
exactly the population it scores in practice. The human annotator labelled
per-SCP compliance given exactly the input the automated judge received ---
the task description, exploit, generated code, relevant SCPs, and DNF formula
--- and the overall verdict is derived from those labels via
\texttt{eval\_dnf} and compared to the judge's verdict.

\begin{center}
\small
\begin{tabular}{lrrr}
\toprule
\textbf{Variant} & \textbf{Prec} & \textbf{Rec} & \textbf{F1} \\
\midrule
2-judge consensus        & 79.8\% & 88.8\% & 84.0 \\
\quad gpt-5-nano alone   & 69.0\% & 86.2\% & 76.7 \\
\quad claude-haiku alone & 74.7\% & 70.0\% & 72.3 \\
\bottomrule
\end{tabular}
\end{center}

The 2-judge consensus outperforms either single LLM on every metric. At the
derived-verdict level, population-reweighted accuracy is 82.3\% with F1\,85.9.
The two automated judges disagreed on 19.5\% of per-SCP labels
($\kappa{=}0.588$), and 54\% of units triggered the third-judge adjudicator,
motivating the multi-annotator design.
\section{Knowledge--Actuation Robustness Check}
\label{app:ka-base}

The main text focuses on the four primary systems. As a robustness check, the table
below expands the knowledge--actuation comparison to ten base models. Actuation is
\securepassatk{}: the joint rate at which generated code is both functional and secure.
The base-model association is \stat{$n=10$, $\rho=0.71$, $p{=}0.025$} for CWEval
and \stat{$n=10$, $\rho=0.63$, $p{=}0.056$} for BaxBench.

\begin{center}
\small
\begin{tabular}{lrrr}
\toprule
\textbf{Model} & \textbf{K} & \textbf{CWEval A} & \textbf{BaxBench A} \\
\midrule
Qwen2.5-14B         & 90.6 & 41.2 & 3.8  \\
SafeCoder           & 90.3 & 40.3 & 4.8  \\
PurpCodeRL          & 88.9 & 30.3 & 4.6  \\
GPT-5.1-Codex-Mini  & 92.3 & 60.5 & 34.2 \\
Claude Haiku 4.5    & 90.1 & 55.5 & 28.1 \\
CodeLlama-13B       & 74.4 & 21.8 & 0.5  \\
GPT-5-Nano          & 90.5 & 58.8 & 34.9 \\
Llama-3.1-8B        & 88.2 & 29.4 & 0.5  \\
Qwen3-30B-Thinking  & 91.7 & 40.3 & 11.5 \\
Qwen3-Coder-30B     & 91.3 & 47.1 & 18.6 \\
\bottomrule
\end{tabular}
\end{center}

\section{Layer-2 Breakdown by Language}
\label{app:language-breakdown}

Tables~\ref{tab:appendix-cweval-language} and~\ref{tab:appendix-baxbench-language} break
the Layer~2 results down by implementation language for the primary systems and methods
from Table~\ref{tab:l2-models}. Each cell reports
\passatk{}/\securepassatk{}/\seccondk{} as percentages. Denominators are fixed within
each benchmark slice; missing generations and missing evaluations count as failures.

\begin{table*}[tp!]
\centering
\caption{CWEval Layer-2 results by language (\passatk{}/\securepassatk{}/\seccondk{}, \%).}
\label{tab:appendix-cweval-language}
\begin{tabular}{llccccc}
\toprule
\textbf{System} & \textbf{Method} &
  \textbf{C} & \textbf{C++} & \textbf{Go} & \textbf{JS} & \textbf{Py} \\
\midrule
Qwen2.5-14B        & Base           & 61/42/68  & 62/29/46  & 53/21/40  & 74/48/65  & 92/60/65 \\
                   & Security-Aware & 68/48/71  & 57/33/58  & 53/21/40  & 65/43/67  & 92/60/65 \\
                   & RCI            & 61/39/63  & 33/29/86  & 21/16/75  & 52/43/83  & 64/52/81 \\
                   & SafeCoder      & 68/48/71  & 43/14/33  & 58/32/55  & 70/43/62  & 84/56/67 \\
                   & PurpCodeRL     & 58/42/72  & 48/24/50  & 53/26/50  & 39/26/67  & 56/28/50 \\
GPT-5.1-Codex-Mini & Base           & 84/65/77  & 86/67/78  & 79/58/73  & 83/57/68  & 92/56/61 \\
                   & Security-Aware & 87/81/93  & 81/67/82  & 74/68/93  & 78/70/89  & 88/72/82 \\
                   & RCI            & 61/58/95  & 67/67/100 & 47/47/100 & 48/48/100 & 72/64/89 \\
                   & Codex Scaffold & 100/94/94 & 100/90/90 & 95/95/100 & 96/91/95  & 92/88/96 \\
\bottomrule
\end{tabular}
\end{table*}

\begin{table*}[tp!]
\centering
\caption{BaxBench Layer-2 results by language (\passatk{}/\securepassatk{}/\seccondk{}, \%).}
\label{tab:appendix-baxbench-language}
\begin{tabular}{llcccccc}
\toprule
\textbf{System} & \textbf{Method} &
  \textbf{Go} & \textbf{JS} & \textbf{PHP} & \textbf{Py} & \textbf{Ruby} & \textbf{Rust} \\
\midrule
Qwen2.5-14B        & Base           & 7/4/50   & 16/4/22  & 0/0/--   & 22/7/32  & 4/0/0    & 0/0/-- \\
                   & Security-Aware & 10/8/88  & 17/5/32  & 0/0/--   & 18/5/30  & 4/4/100  & 0/0/-- \\
                   & RCI            & 19/10/50 & 16/9/56  & 0/0/--   & 16/9/56  & 4/4/100  & 0/0/-- \\
                   & SafeCoder      & 15/8/54  & 21/9/43  & 0/0/--   & 28/14/52 & 0/0/--   & 0/0/-- \\
                   & PurpCodeRL     & 13/6/45  & 13/4/33  & 0/0/--   & 12/7/62  & 0/0/--   & 0/0/-- \\
GPT-5.1-Codex-Mini & Base           & 60/36/60 & 54/38/70 & 36/21/60 & 68/42/62 & 32/29/89 & 0/0/-- \\
                   & Security-Aware & 71/51/72 & 52/36/69 & 50/36/71 & 61/44/72 & 39/29/73 & 0/0/-- \\
                   & RCI            & 40/32/79 & 52/41/79 & 14/11/75 & 43/34/79 & 25/21/86 & 0/0/-- \\
                   & Codex Scaffold & 75/51/68 & 69/49/71 & 43/32/75 & 68/51/75 & 0/0/--   & 0/0/-- \\
\bottomrule
\end{tabular}
\end{table*}

\end{document}